# Evaluation of Copper, Aluminum and Nickel Interatomic Potentials on Predicting the Elastic Properties

Seyed Moein Rassoulinejad-Mousavi, Yijin Mao and Yuwen Zhang[1]

*Department of Mechanical and Aerospace Engineering, University of Missouri, Columbia, Missouri, 65211, USA*

**Abstract**

Choice of appropriate force field is one of the main concerns of any atomistic simulation that needs to be seriously considered in order to yield reliable results. Since, investigations on mechanical behavior of materials at micro/nanoscale has been becoming much more widespread, it is necessary to determine an adequate potential which accurately models the interaction of the atoms for desired applications. In this framework, reliability of multiple embedded atom method based interatomic potentials for predicting the elastic properties was investigated. Assessments were carried out for different copper, aluminum and nickel interatomic potentials at room temperature which is considered as the most applicable case. Examined force fields for the three species were taken from online repositories of National Institute of Standards and Technology (NIST), as well as the Sandia National Laboratories, the LAMMPS database. Using molecular dynamic simulations, the three independent elastic constants, $C_{11}$, $C_{12}$ and $C_{44}$ were found for *Cu*, *Al* and *Ni* cubic single crystals. Voigt-Reuss-Hill approximation was then implemented to convert elastic constants of the single crystals into isotropic polycrystalline elastic moduli including Bulk, Shear and Young's modulus as well as Poisson's ratio. Simulation results from massive molecular dynamic were compared with available experimental data in the literature to justify the robustness of each potential for each species. Eventually, accurate interatomic potentials have been recommended for finding each of the elastic properties of the pure species. Exactitude of the elastic properties was found to be sensitive to the choice of the force fields. Those potentials were fitted for a specific compound may not necessarily work accurately for all the existing pure species. Tabulated results in this paper might be used as a benchmark to increase assurance of using the interatomic potential that was designated for a problem.

**Introduction**

There has been a growing of interest in understanding the mechanical properties of nanosized structures, such as nanowires, nanofilms, nanoplates, nanotubes, and nanocrystals, due to their emergence in various area of science and technology such as nanoelectromechanical systems (NEMS) [1-3]. Investigations regarding the elastic properties of nanocomponents, are playing important roles in determining the NEMS sensitivity and speed of response. Therefore, many experimental, computational and theoretical studies in this field have been intensively reported in the recent years. Experimentally, Cuenot *et al.* [4] studied surface tension effect on the mechanical properties of lead nanowires and polypyrrole nanotubes using atomic force microscopy. They found that decreasing the diameter of the nanomaterial led to an increase in modulus of elasticity. Fonseca et al.[5] proposed a model to predict the elastic properties of the composite material of an amorphous nanowire. They presented experimental setup for two scenarios: a) static configuration in which the nanowire was maintained twisted, and b) the twisting and tensioning of a nanowire. Peng et al.[6] experimentally studied size-dependent mechanical properties of single crystalline nickel nanowires with different diameters from 100 to 300 nm and crystalline orientations using in-situ tensile tests in a scanning electron microscope. Their important observations were that critical resolved shear stress strongly depended on the sample's crystalline size and was independent of its orientation. Using the same method, Li et al.[7] obtained mechanical properties of individual InAs nanowires, synthesized by metal organic chemical vapor deposition and molecular beam epitaxy, and determined that Young's modulus had no dependency on nanowire diameter. Computationally, Liang and Upmanyu[8] studied the size-dependent elasticity of copper nanowires using molecular statics approach based on embedded atom method (EAM) potential developed by Mishin et al.[9]. They concluded that the Young's modulus of the nanowire was lower than that of bulk copper strained to an equivalent compressive strain. They also found that the nanowire surface was always softer than the (equivalent) bulk. Yuan and Huang[10] investigated the size-dependent elasticity of amorphous silica nanowire using molecular dynamics simulation (MDS). Their results showed that method of sample preparation significantly affected the elastic response of silica nanowires. Moreover, they concluded that overall elasticity could be controlled through the coupling between the surface stiffening and the core softening. Liu et al.[11] used atomistic simulations to investigate the role of the surface on the size-dependent mechanical properties of copper nanowire with diameters ranging from 2 to 20 nm under tensile load. Their molecular dynamics results verified that the elastic properties of the nanowire was directly associated to the proportion of surface to bulk type atoms. Zanjani and Lukes[12] investigated the size effect on Young's modulus, bulk modulus, and Poisson's ratios of *CdSe* nanocrystal superlattices using fully atomistic MDS, coarse-grained models, and effective medium theory. They found that the fully atomistic models were the most reliable method because of its good agreement with previously reported results in the literature dealing with a similar system. They also concluded that with increasing nanocrystal core size, Young's modulus and bulk modulus increased while it led to a decrease in Poisson's ratio. Based on a developed modified core-shell model, Hai-Yan et al.[13] investigated the effect of surface elasticity on Young's modulus to the bending of nanobeams. They found significant size-dependent effect of elastic modulus with characteristic size reduction, especially below 100 nm. Wang et al.[14] studied the dependency of Young's modulus to size and temperature for a silicon nanoplate based on a semi-continuum approach. Their results showed that Young's modulus decreased while temperature increased. They also found that thickness of nanoplate influenced the elastic property of the silicon nanoplate with its thickness less than 10 nm and remained constant when it was thicker. Further theoretical, numerical and experimental investigations regarding mechanical properties of nanostructures can be seen in the references[15-18].

---

[1] Corresponding author. Email: zhangyu@missouri.edu.

    Molecular dynamics simulation (MDS) has become of major interest for researchers as a powerful modeling tool in solving various micro/nanoscale problems in number of fields such as biophysics, biochemistry, material science, engineering and structural biology[19-23]. Among different potentials that can be applied for MDS, the embedded-atom method (EAM)[24] is a widely used pair-wise interatomic potential for metallic systems and their alloys. Initial form of EAM potential introduced by [24] had certain limits in predicting some physical phenomena, and afterward some improvements were made by scientists for different purposes. Therefore, many EAM

interatomic potentials for different pure species and their alloys are generated based on different fitting criteria in the literature and most of them can be found in interatomic potentials repository of NIST [25] and the LAMMPS database from Sandia National Laboratories [26]. However, it is aware that limited works has been done in the recent years to show the significance of implementing an appropriate force field, also known as interatomic potentials, for MDS in conducting an accurate simulation. Suggestions regarding this topic has been released by Becker *et al.* [27] via considering the significance of appropriate choice of force field in materials science and engineering through giving some examples and investigation of the elastic constant for aluminum, using some of the *Al* interatomic potentials managed by the NIST Interatomic Potentials Repository (IPR).

Trautt et al. [28] discussed the automation of stacking fault energy calculations and their application to additional elements by explaining how the calculations can be modified using the Python script. They compared the fault energy obtained by different interatomic potentials of different elements with the results of the density functional theory. Kalidindi et al. [29] considered the application of data science tools to quantify and to distinguish between structures and models in molecular dynamics datasets. They examined the *Al* potentials in the IPR to the variation of the principal component scores as a function of temperature for the different force fields. Their results also indicated the utility and the viability of utilizing rigorous structure quantification protocols to results predicted by MD.

Since each interatomic potential is generally based on a specific problem and fit to reference data within a certain range of composition, temperature, structure. Outside of that range, the interatomic potential may not provide physically meaningful results. In the other words, it is very difficult to develop a universal interatomic potential that works appropriately for all applications. On the other hand, the robustness, accuracy and validity of an atomistic simulations hinge on the appropriate choice of force fields. Hence, it is necessary to make sure that the performance of an interatomic potential is adequate for a specific purpose with desired condition. To serve this objective and make a shortcut for users to find their best force fields, meanwhile increasing their assurance, among so many available interatomic potential, a number of force fields are examined in this work for the three popular metallic materials *Cu*, *Al* and *Ni* which are popular in the micro/nanoscale science and technologies. Room temperature which is the most applicable and realistic case to test the materials' properties is considered in the current simulations. Using molecular dynamic simulation, elastic properties of the cubical single crystals will be computed based on the stress-strain curves by imposing uniaxial tensile as well as shear strains to the system. The predicted elastic constants are compared with experimental values with the purpose of determining which potentials are accurate for the interested metals. Eventually, based on the Voigt-Reuss-Hill (VRH) approximation [30] single crystals' elastic constants are converted into isotropic elastic moduli thereafter those values can be meaningfully compared with the experimental values for Copper, Aluminum and Nickel.

**Computational Details**

Regarding with the materials mechanical properties, this should be noted that the single-crystal elasticity will not be isotropic in general. The number of independent material parameters depends on the level of symmetry of the crystal structure. In fact, the general form of the Hooke's law for the 36 elastic constants is given by the following stiffness tensor,

$$\underline{C} = \begin{bmatrix} C_{11} & C_{12} & C_{13} & C_{14} & C_{15} & C_{16} \\ C_{21} & C_{22} & C_{23} & C_{24} & C_{25} & C_{26} \\ C_{31} & C_{32} & C_{33} & C_{34} & C_{35} & C_{36} \\ C_{41} & C_{42} & C_{43} & C_{44} & C_{45} & C_{46} \\ C_{51} & C_{52} & C_{53} & C_{54} & C_{55} & C_{56} \\ C_{61} & C_{62} & C_{63} & C_{64} & C_{65} & C_{66} \end{bmatrix} \qquad (1)$$

For the cubic structure, the elasticity matrix can be written in terms of three independent elastic stiffness constants since $C_{11} = C_{22} = C_{33}$, $C_{12} = C_{21} = C_{23} = C_{32} = C_{13} = C_{31}$ and $C_{44} = C_{55} = C_{66}$ because x-, y-and z- axes are identical by symmetry. In addition, the off diagonal shear components are zero which means $C_{45} = C_{54} = C_{56} = C_{65} = C_{46} = C_{64} = 0$ [31,32]. So the stiffness tensor can be simplified into a three variable independent form as,

$$\underline{C} = \begin{bmatrix} C_{11} & C_{12} & C_{12} & 0 & 0 & 0 \\ C_{12} & C_{11} & C_{12} & 0 & 0 & 0 \\ C_{12} & C_{12} & C_{11} & 0 & 0 & 0 \\ 0 & 0 & 0 & C_{44} & 0 & 0 \\ 0 & 0 & 0 & 0 & C_{44} & 0 \\ 0 & 0 & 0 & 0 & 0 & C_{44} \end{bmatrix} \qquad (2)$$

So, for a cubic single crystal three independent elastic constants $C_{11}$, $C_{12}$ and $C_{44}$ are going to be found using each interatomic potential.

Elastic constants $C$ relate the strain $\varepsilon$ and the stress $\sigma$ in a linear fashion:

$$\sigma_{ij} = \sum_{kl} C_{ijkl} \varepsilon_{kl}, \qquad (3)$$

which relates to elastic energy density $U$ that is defined as follows [33] for a cubic crystal x- along [100], y || [010], z || [001]:

$$U = \frac{1}{2} C_{11}(e_{xx}^2 + e_{yy}^2 + e_{zz}^2) + \frac{1}{2} C_{44}(e_{yz}^2 + e_{zx}^2 + e_{xy}^2) + C_{12}(e_{yy}e_{zz} + e_{zz}e_{xx} + e_{xx}e_{yy}), \qquad (4)$$

where $e_{ij}$ are strain components by the relations:

$$e_{xx} \equiv \varepsilon_{xx} = \frac{\partial u}{\partial x}; \qquad e_{yy} \equiv \varepsilon_{yy} = \frac{\partial v}{\partial y}; \qquad e_{xx} \equiv \varepsilon_{zz} = \frac{\partial w}{\partial z},$$

$$e_{xy} \equiv \varepsilon_{yx} + \varepsilon_{xy} = \frac{\partial u}{\partial y} + \frac{\partial v}{\partial x}; \quad e_{yz} \equiv \varepsilon_{zy} + \varepsilon_{yz} = \frac{\partial v}{\partial z} + \frac{\partial w}{\partial y}; \quad e_{yz} \equiv \varepsilon_{zy} + \varepsilon_{yz} = \frac{\partial u}{\partial z} + \frac{\partial w}{\partial x}; \qquad (5)$$

where $u, v$ and $w$ are displacements in x-, y- and z- directions.

Further, from (4),

$$\frac{\partial U}{\partial e_{ij}} = \sigma_{ij} \tag{6}$$

Therefore, we derive the followings according to (3) or (6):

$$\sigma_{xx} = C_{11}e_{xx} + C_{12}(e_{yy} + e_{zz}), \tag{7}$$

$$\sigma_{yy} = C_{11}e_{yy} + C_{12}(e_{xx} + e_{zz}), \tag{8}$$

$$\sigma_{zz} = C_{11}e_{zz} + C_{12}(e_{xx} + e_{yy}), \tag{9}$$

$$\sigma_{xy} = C_{44}\varepsilon_{xy} + C_{44}\varepsilon_{yx} = C_{44}e_{xy}. \tag{10}$$

where, $e_{xy}$ is also known as $\gamma_{xy}$, which is the engineering shear strain.

On the simulations to find the three independent elastic constants, Large-scale atomic/molecular massively parallel simulator (LAMMPS), a classical molecular dynamics solver, is used for the current work [34]. A cubic box with size of $70a \times 70a \times 70a$ is created for copper, aluminum and nickel with lattice constant $a$. The periodic boundary condition is applied in the x-, y- and z-directions to find the bulk properties for comparison with experimental data in the literature. The materials being considered are pure substance of solid *Cu*, *Al* and *Ni* arranged in the face centered cubic (FCC) fashion. Different styles of the EAM potential is adopted in the simulations. According to Daw and Baskes [24], the potential energy of an embedded atom *i* can be approximated as,

$$E_{tot} = \sum_i F_i(\rho_{h,i}) + \frac{1}{2}\sum_i \sum_{j(\neq i)} \phi_{ij}(R_{ij}) \tag{11}$$

where $F_i(\rho)$ is the embedding energy for embedding atom *i* into the host electron density $\rho$, and $\phi_{ij}(R_{ij})$ is the pair potential which is a function of the distance $R$ between atoms *i* and *j*. The $\rho_{h,i}$ represents the host electron density at atom *i* due to the remaining atoms of the system which is approximated by the superposition of atomic densities as follows,

$$\rho_{h,i} = \sum_{j(\neq i)} \rho_j^a(R_{ij}) \tag{12}$$

where $\rho_j^a(R)$ is the electron density at the site of atom *i* due to the presence of atom *j* at a distance of $R$.

The potential functions in (11) and (12) are usually treated as some fitting functions which are proposed by some researchers in consideration of the physical properties of the interested metals as well as their alloys. Generally, the EAM potential is simple, however its embedded energy and pair potential are given in the form of spline functions which leads to some inconvenience for calculations [35]. This explains why there are so many EAM based interatomic potentials developed or optimized in the literature. In this paper, the fitted functions of EAM based potentials generated for a specific compound or a pure species are examined to obtain elastic constants of a cubic single crystal and different elastic moduli for the three elements. To achieve this, the single crystal *Cu*, *Al* and *Ni* lattices need to be deformed under a uniaxial and a shear strain, independently.

Finding elastic constants at finite temperature is a little tricky and more complicated with respect to zero-temperature properties. This is why users need to pay more attention to the system equilibration, constrains and methods of calculations. Here, the simulation procedure for finding the three independent elastic constants of cubical single crystals is explained in detail.

First, an instantaneous temperature of 500K was defined in terms of the instantaneous kinetic energy to generate an ensemble of velocities using a random number generator with the specified seed as the specified temperature. Then, the entire system was equilibrated at a uniform temperature of 500K under NVT dynamic, for 10 ps. Time step $\Delta t = 1 fs$ is used throughout the simulations. After that, the relaxation was continued with the same temperature for 25 ps using isothermal-isobaric (NPT) ensemble. In the next stage of relaxation, the structure was cooled down to the desired temperature of 300 K for 25 ps followed by a duration of 25 ps at the same temperature. It is worth to note that, in all equilibration stages, linear momentum was zeroed by subtracting the center-of-mass velocity of the group from each atom. To calculate the anisotropic elastic stiffness constants, $C_{11}$ and $C_{12}$, a simple uniaxial tensile strain rate of $10^{-3}$ ps$^{-1}$ (strain increases 0.1% every picosecond) was applied along the [100] (x- direction) at 300 K which leads to nonzero stress components $\sigma_{xx}, \sigma_{yy}, \sigma_{zz}$ [36]. The strains in the y- and z- directions were both controlled to be zero under the NVT ensemble to find the two stiffness constants based on (7) and (8) having $\varepsilon_{yy} = \varepsilon_{zz} = 0$. Once the stress-strain curves are obtained, it is straightforward to find the $C_{11}$ and $C_{12}$ from the slope of the linear part of $\sigma_{xx}$ and $\sigma_{yy}$ versus strain, respectively. Figure 1 displays a schematic of the two created atomistic systems under the applied uniaxial tensile and shear strains.

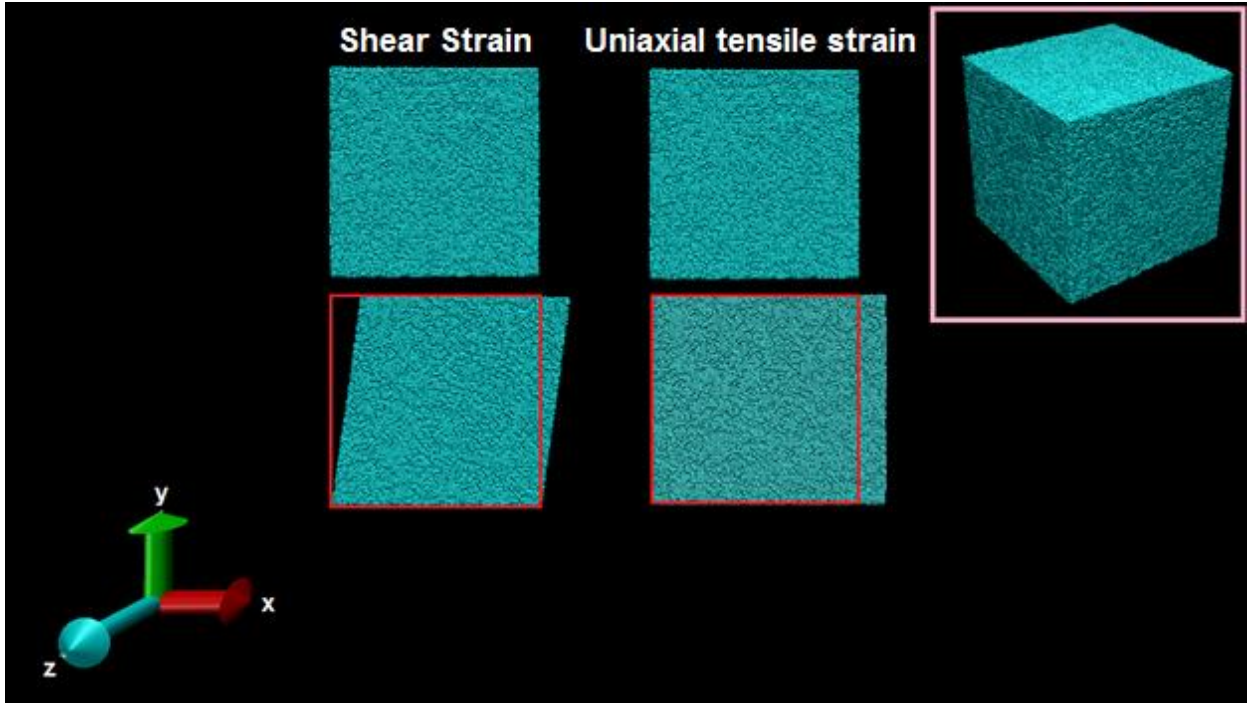

Fig. 1 Schematic of applied uniaxial and shear strains to the atomistic systems

To find $C_{44}$, a prism region was created rather than a block that was used for the uniaxial strain, to define a triclinic simulation box with initial tilt factors of zero. Then, with the same equilibration steps and under the NVT ensemble, the system was distorted in the [110] direction by applying an engineering shear strain rate of $10^{-3}$ ps$^{-1}$ so that the crystal is no longer cuboidal. Afterward, $C_{44}$ has been calculated by finding the slope of linear portion of yield curve for $\sigma_{xy}$ versus the associated strain $e_{xy}$ according to (10).

Once the elastic constants using different potentials are found, using the Viogt-Reuss-Hill approximation [30] which is an averaging scheme, the anisotropic single-crystal elastic constants can be converted into isotropic polycrystalline elastic moduli. For a single-phase crystalline aggregate made of crystals that are slightly anisotropic, the approximation gives the realistic values of isotropic elastic moduli [37]. This approach combines the upper and lower bounds by assuming the average of values obtained through the Voigt [38] and Reuss [39] averaging methods. In the upper bound (Voigt) the strain assumes to be uniform and continuous whereas the stresses are allowed to be discontinuous. In the lower bound (Reuss) the stresses are assumed to be continuous and the strains can be discontinuous.

Bulk modulus is defined similarly under all the three approximations and can be exactly defined as follows,

$$B_{VRH} = B_V = B_R = \frac{C_{11} + 2C_{12}}{3}, \tag{13}$$

and the upper and lower bounds of shear moduli are,

$$G_V = \frac{C_{11} - C_{12} + 3C_{44}}{5}, \tag{14}$$

and

$$G_R = \frac{5C_{44}(C_{11} - C_{12})}{4C_{44} + 3(C_{11} - C_{12})}, \tag{15}$$

Knowing shear modulus $G$ and Bulk modulus $B$, it is then straightforward to find Young's modulus $E$ and Poisson's ratio $\nu$, from the well-known isotropic relations as follows,

$$E = \frac{9GB}{3B + G}, \tag{16}$$

and

$$\nu = \frac{3B - 2G}{2(3B + G)}, \tag{17}$$

Eventually, based on the Hill empirical average [30], we have the isotropic elastic modulus as follows,

$$G_{VRH} = \frac{G_V + G_R}{2}, \tag{18}$$

$$E_{VRH} = \frac{E_V + E_R}{2}, \tag{19}$$

$$\nu_{VRH} = \frac{\nu_V + \nu_R}{2}. \tag{20}$$

**Results and Discussions**

Results, using different classes of potentials, eam, eam/alloy, Finnis-Sinclair [40], for three independents elastic stiffness constants of $Cu$, $Al$ and $Ni$ single crystals are tabulated through Tables 1 to 3. Elastic constants $C_{11}$, $C_{12}$ and $C_{44}$ were found from the slope of the elastic regime of the stress-strain curves, where the growth of stress versus strain is linear. These elastic stiffness constants were obtained using a curve fitting technique by defining a stipulation that in linear regression, the $R^2$ coefficient of determination should be greater than 0.999 to achieve an accurate judgment on the results as much as possible. This is the reason that a large simulation box with more than a million atoms was created to minimize the temperature deviation due to thermal velocity fluctuations and thereafter yield a smooth linear stress versus strain relation rather than an oscillatory one. Figure 2 shows the effect of number of atoms on simulation temperature through the applied strains. As can be seen, increasing the number of atoms decreases the deviations from desired temperature which leads to a more smooth line for stress-strain curve and eventually a better and more reliable linear fit. Furthermore, it is obvious from the Fig.2 that deviation from 300K is about 40K for 500 atoms while it is less than 1K for more than one million atoms, which completely makes sense from statistical mechanics point of view.

Table 1 Accuracy of different Copper interatomic potentials with respect to experimental data for $C_{11}$=168.4 GPa, $C_{12}$=121.4 GPa and $C_{44}$=75.4 Gpa [45]

| Copper Potentials | $C_{11}$ | Error | $C_{12}$ | Error | $C_{44}$ | Error |
|---|---|---|---|---|---|---|
| Mendelev_Cu2_2012.eam.fs [52] | 164.56 | 2.28% | 121.31 | 0.07% | 78.97 | 4.73% |
| Cu1.eam.fs [52] | 160.44 | 4.73 | 120.50 | 0.74 | 79.24 | 5.09 |
| Cu-Zr.eam.fs [53] | 159.72 | 5.15 | 120.70 | 0.58 | 78.53 | 4.15 |
| Cu-Zr_2.eam.fs [54] | 159.7 | 5.17 | 120.72 | 0.56 | 78.49 | 4.09 |
| Cu_mishin1.eam.alloy [9] | 161.61 | 4.03 | 117.14 | 3.51 | 71.15 | 5.63 |
| Cu01.eam.alloy [9] | 161.1 | 4.33 | 116.62 | 3.94 | 71.38 | 5.33 |
| FeCuNi.eam.alloy [43] | 161.25 | 4.25 | 116.55 | 4.00 | 71.51 | 5.15 |
| CuAg.eam.alloy [55] | 161.23 | 4.26 | 116.38 | 4.14 | 71.14 | 5.65 |
| Cu_u6.eam [56] | 150.83 | 10.43 | 112.2 | 7.58 | 70.42 | 6.60 |
| Cu_smf7.eam [57] | 150.45 | 10.66 | 111.25 | 8.36 | 71.65 | 4.97 |
| Cu_u3.eam [58] | 150.07 | 10.88 | 112.13 | 7.64 | 68.72 | 8.86 |
| PbCu.setfl [59] | 149.74 | 11.08 | 111.85 | 7.87 | 68.7 | 8.88 |
| Cu_zhou.eam.alloy [60] | 161.1 | 4.33 | 106.07 | 12.63 | 64.91 | 13.91 |
| CuNi.eam.alloy [61] | 148.33 | 11.92 | 112.4 | 7.41 | 76.18 | 1.03 |
| AlCu.eam.alloy [35] | 147.83 | 12.21 | 111.24 | 8.37 | 66.62 | 11.64 |
| Cu.set [62] | 138.51 | 17.75 | 99.8 | 17.79 | 54.26 | 28.03 |
| Cu_Ag_ymwu.eam.alloy [63] | 116.1 | 31.06 | 87.43 | 27.98 | 47.13 | 37.49 |

Table 2 Accuracy of different Aluminum interatomic potentials with respect to experimental data for $C_{11}$=107.3 GPa, $C_{12}$=60.08 GPa and $C_{44}$=28.3 GPa [47]

| Aluminum Potentials | $C_{11}$ | Error | $C_{12}$ | Error | $C_{44}$ | Error |
|---|---|---|---|---|---|---|
| Mishin-Ni-Al-Co-2013.eam.alloy[41] | 107.21 | 0.08% | 60.60 | 0.33% | 32.88 | 16.18% |
| Al99.eam.alloy [64] | 107.03 | 0.25 | 61.06 | 0.43 | 31.05 | 9.72 |
| Mishin-NiAl2009 [42] | 107.5 | 0.19 | 61.25 | 0.74 | 33.2 | 17.31 |
| Mishin-Al-Co-2013.eam.alloy [41] | 107.58 | 0.26 | 61.82 | 1.68 | 32.88 | 16.18 |
| Mg-Al-set.eam.alloy [65] | 107.08 | 0.21 | 58.35 | 4.03 | 31.6 | 11.66 |
| Al-Mg.eam.fs [66] | 103.86 | 3.21 | 63.8 | 4.93 | 30.48 | 7.70 |
| AlPb-setfl.eam.alloy [67] | 105.63 | 1.56 | 58.2 | 4.28 | 31.85 | 12.54 |
| Zope-Ti-Al-2003.eam.alloy [68] | 98.27 | 8.42 | 56.23 | 7.52 | 26.02 | 8.06 |
| Al03.eam.alloy [68] | 98.2 | 8.48 | 55.88 | 8.09 | 26.42 | 6.64 |
| NiAl.eam.alloy [69] | 100.52 | 6.32 | 54.65 | 10.12 | 26.14 | 7.63 |
| AlFe-mm [26] | 95.25 | 11.23 | 55.82 | 8.19 | 30.28 | 7.00 |
| Al-Fe.eam.fs [70] | 94.15 | 12.26 | 55.18 | 9.24 | 30.28 | 7.00 |
| Almm [25] | 103.95 | 3.12 | 63.96 | 5.20 | 36.7 | 29.68 |
| Al1.eam.fs [52] | 101.53 | 5.38 | 64.5 | 6.09 | 36.67 | 29.58 |
| Al_jnp.eam [50] | 96.58 | 9.99 | 74.61 | 22.71 | 40.43 | 42.86 |
| Al-LEA.eam.alloy [71] | 123.09 | 14.72 | 64.18 | 5.56 | 39.5 | 39.58 |
| AlO.eam.alloy [51] | 82.92 | 22.72 | 69.3 | 13.98 | 30.37 | 7.31 |
| Al_zhou.eam.alloy [60] | 81.38 | 24.16 | 58.12 | 4.41 | 20.15 | 28.80 |
| CoAl.eam.alloy [72] | 83.16 | 22.50 | 55.1 | 9.37 | 21.41 | 24.35 |
| AlCu.eam.alloy [35] | 73.97 | 31.06 | 59.24 | 2.57 | 51.1 | 80.57 |
| Farkas_Nb-Ti-Al_1996.eam.alloy [73] | 171.64 | 59.96 | 116.1 | 93.24 | 91.83 | 224.49 |
| Al.set [62] | 61.16 | 43.00 | 45.74 | 24.77 | 10.86 | 61.63 |
| NiAlH_jea.eam.fs [74] | 58.26 | 45.70 | 33.5 | 44.90 | 58.26 | 105.87 |
| NiAlH_jea.eam.alloy [74] | 58.98 | 45.03 | 32.9 | 45.89 | 58.98 | 108.41 |

Table 3 Accuracy of different nickel interatomic potentials with respect to experimental data for
$C_{11}$=253 GPa, $C_{12}$=152 GPa and $C_{44}$= 124Gpa[49]

| Nickel Potentials | $C_{11}$ | Error | $C_{12}$ | Error | $C_{44}$ | Error |
|---|---|---|---|---|---|---|
| Ni.eam.fs [75] | 245.13 | 1.95% | 147.58 | 1.61% | 132.33 | 6.718% |
| FeNiCr_Bonny_2013_ptDef.eam.alloy [76] | 248.14 | 0.74 | 144.83 | 3.45 | 116.75 | 5.847 |
| Mishin-Ni-Al-2009.eam.alloy [42] | 224.76 | 10.10 | 132.14 | 11.91 | 117.48 | 5.258 |
| CuNi.eam.alloy [61] | 223.57 | 10.57 | 135.87 | 9.42 | 113.25 | 8.669 |
| NiAl.eam.alloy [69] | 220.2 | 11.92 | 137.32 | 8.45 | 118.89 | 4.121 |
| Mishin-Ni-Al-Co-2013.eam.alloy [41] | 221.1 | 11.56 | 138.21 | 7.86 | 109.65 | 11.573 |
| Mishin-Ni-Co-2013.eam.alloy [41] | 219.55 | 12.18 | 138.93 | 7.38 | 109.5 | 11.694 |
| Ni99.eam.alloy [64] | 283.7 | 13.48 | 167.03 | 11.35 | 135.45 | 9.234 |
| Ni_smf7.eam [57] | 213 | 14.80 | 143.78 | 4.15 | 114.58 | 7.597 |
| Ni_u6.eam [56] | 211.06 | 15.58 | 143.81 | 4.13 | 122.78 | 0.984 |
| Ni_u3.eam [58] | 210.98 | 15.61 | 143.04 | 4.64 | 117.02 | 5.629 |
| FeCuNi.eam.alloy [43] | 219.66 | 12.14 | 138.08 | 7.95 | 102.42 | 17.403 |
| Fe-Ni.eam.alloy [77] | 219.89 | 12.04 | 137.42 | 8.39 | 98.28 | 20.742 |
| FeNiCr.eam.alloy [78] | 217.87 | 12.85 | 131.5 | 12.33 | 107.04 | 13.677 |
| Ni.set [62] | 209.34 | 16.26 | 127.96 | 14.69 | 109.4 | 11.774 |
| NiAlH_jea.eam.alloy [74] | 207.61 | 16.96 | 127.57 | 14.95 | 106.21 | 14.347 |
| Ni-Zr_Mendelev_2014.eam.fs [79] | 206.32 | 17.47 | 126.98 | 15.35 | 114.15 | 7.944 |
| Ni1_Mendelev_2012.eam.fs [80] | 206.73 | 17.31 | 125.95 | 16.03 | 118.93 | 4.089 |

Simulation results based on each interatomic potential have been compared with experimental data available in the literature at temperature of 300K and meanwhile the corresponding relative errors are reported. It should be noted, that one may find less/more relative errors with respect to experimental values by using a different criterion, such as using R-squared value less than 0.999.

As seen from the Tables 1 to 3, most of the examined potentials are yielding acceptable results in predicting all the three elastic constants according to their relative errors. Tables 1 to 3 also exemplify that some of the interatomic potentials are capable for accurate determination of one or two elastic constant rather than all, and may fail for the other one/s. This is the reason that user must be very careful about whether the force field works well for their interested problem or not. As an example in Table 2, *Mishin-Ni-Al-Co-2013.eam.alloy* [41] and *Mishin-NiAl2009* [42] predict the $C_{11}$ and $C_{12}$ for the Aluminum with relative errors less than 1% while yield 16.18% and 17.31% error for $C_{44}$, respectively. On the contrary, *Ni1_Mendelev_2012.eam.fs* predicts nickel $C_{11}$ and $C_{12}$ with about 20.00% error while yields the $C_{44}$ with 3.42% error. In the other scenario a potential like *Ni_u6.eam* gives 4.30% and 0.99% error for $C_{12}$ and $C_{44}$ respectively but predict $C_{11}$ with 18.45% error from experimental result. Furthermore, among the potentials that have been created for a specific compound, they may work well for one or some of the elements but poorly predict the elastic properties for the others. For instance, *FeCuNi.eam.alloy* [43] predict the three elastic stiffness constants with relative error less than 6.00% for copper but yields 12.14% error for $C_{11}$, 7.95% for $C_{12}$ and 17.40% for $C_{44}$ for nickel. These inaccuracies of potentials which are not providing reasonable results could be due to the limit of the fitting range of composition or structure for each interatomic potential. For this reason, it may lead to no physically meaningful results outside of that range.

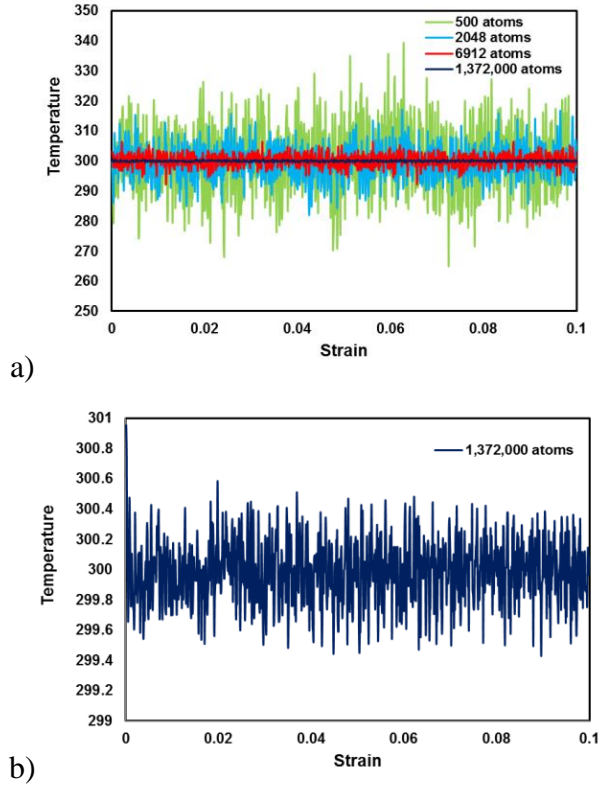

Fig. 2 a) Effect of number of atoms on deviation of temperature from desired one. b) uniform temperature near to 300K with more than one million atoms

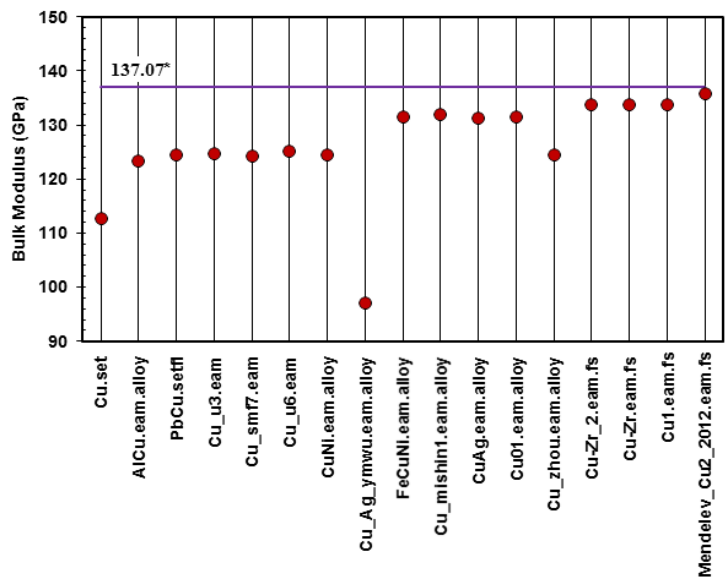

Fig. 3 Copper Bulk modulus predicted by different force fields

Afterward, single crystal elastic stiffness constants have been converted into isotropic elastic moduli including Bulk, shear and Young's modulus as well as Poisson's ratio in Figures 3-14 using VRH approximation for different interatomic potentials listed in Tables 1-3. Simulation results are compared with experimental ones for *Cu* [44,45], *Al* [46,47] and *Ni* [48,49] for these interested mechanical properties.

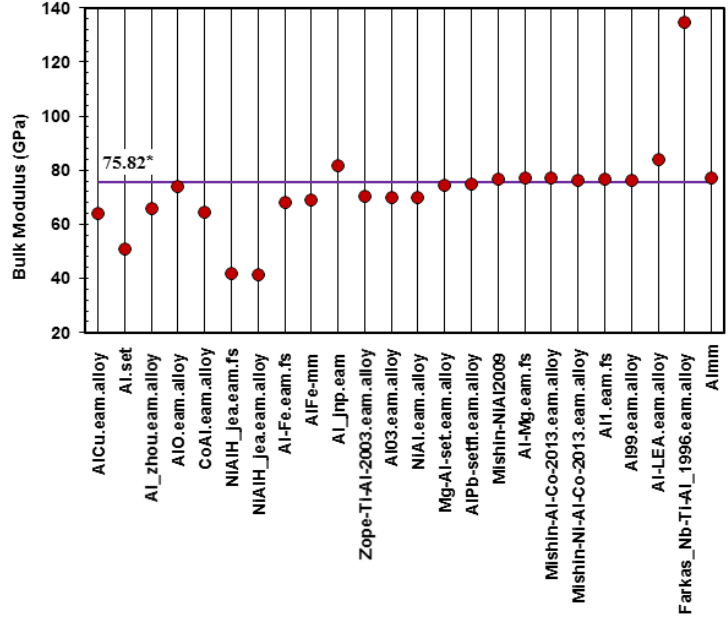

Fig. 4 Aluminum Bulk modulus predicted by different force fields

Figure 3-5 show the Bulk modulus of the three studied species obtained by each elemental potential and presents the deviation of MDS results from experimental data. According to Fig.3, among examined elemental potentials, copper bulk modulus can be accurately obtained using *Cu-Zr_2.eam.fs, Cu-Zr.eam.fs, Cu1.eam.fs and Mendelev_Cu2_2012.eam.fs* with less than 3.00% relative error from experimental value and also *FeCuNi.eam.alloy, Cu_mishin1.eam.alloy, CuAg.eam.alloy, Cu01.eam.alloy* with less than 4.50%. As can be seen from (13), bulk modulus depends only on the two independent elastic constant $C_{11}$ and $C_{12}$. It is found from Fig.3, potentials that successfully predicted these two elastic constants shows their capability in predicting Bulk modulus accurately as well. However, on the other side a question arises according to Fig.4 that why some potentials fail to predict both $C_{11}$ and $C_{12}$ accurately, but they succeed in predicting Bulk modulus? How logical and reliable they are? The answer to this question is that, while bulk modulus is the result of an algebraic expression, two inaccurate variables can also lead to a desirable value. It is absolutely fortuitous and unreliable potential to account on that for users' simulation in considering the present problem. Sensible instances of this kind of deceptive observations for bulk modulus results are *Al_jnp.eam* [50] and *AlO.eam.alloy* [51] for Aluminum according to Fig.4 and Table 2. This can sparkle a debate on the way of finding bulk modulus. So, this can be concluded that the potentials: *Mishin-Ni-Al-Co-2013.eam.alloy, Al99.eam.alloy, Mishin-NiAl2009, Mishin-Al-Co-2013.eam.alloy, Mg-Al-set.eam.alloy, Al-Mg.eam.fs, AlPb-setfl.eam.alloy, Almm* and *Al1.eam.fs,* might be suggested to the users for finding aluminum bulk modulus as they predicted this property with less than 2.00% relative error, and also the two related

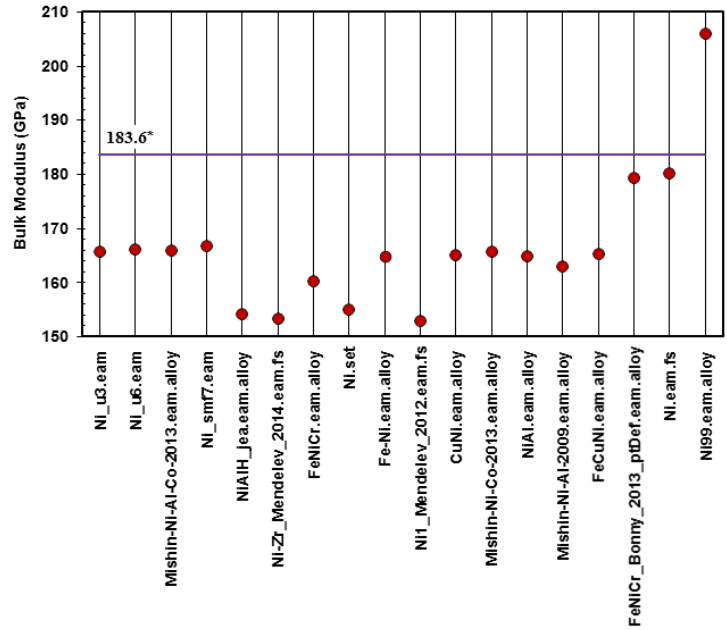

Fig. 5 Nickle Bulk modulus predicted by different force fields

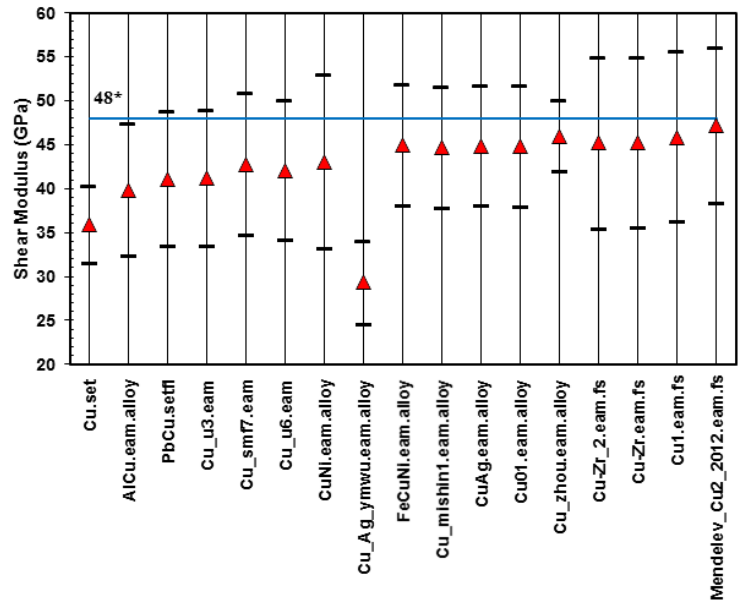

Fig. 6 Copper shear modulus predicted by VRH approximation using different force fields

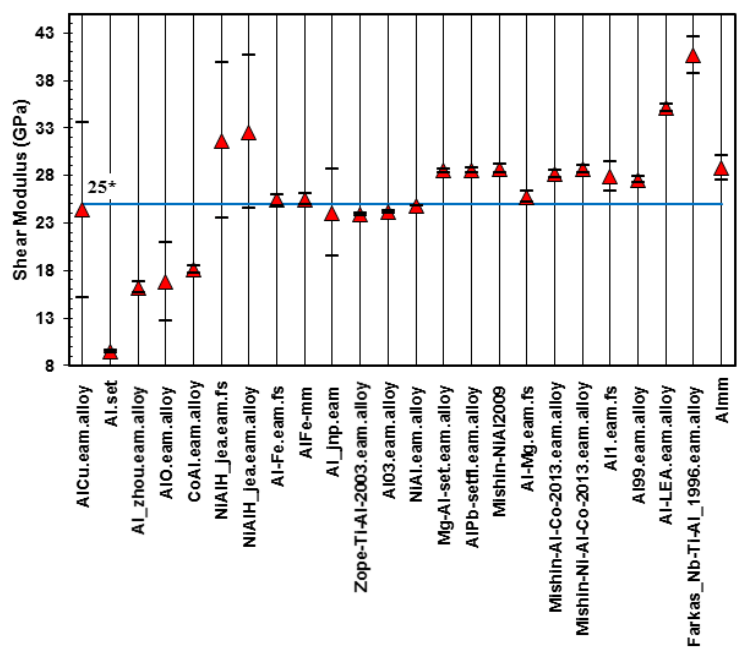

Fig. 7 Aluminum Shear modulus predicted by VRH approximation using different force fields

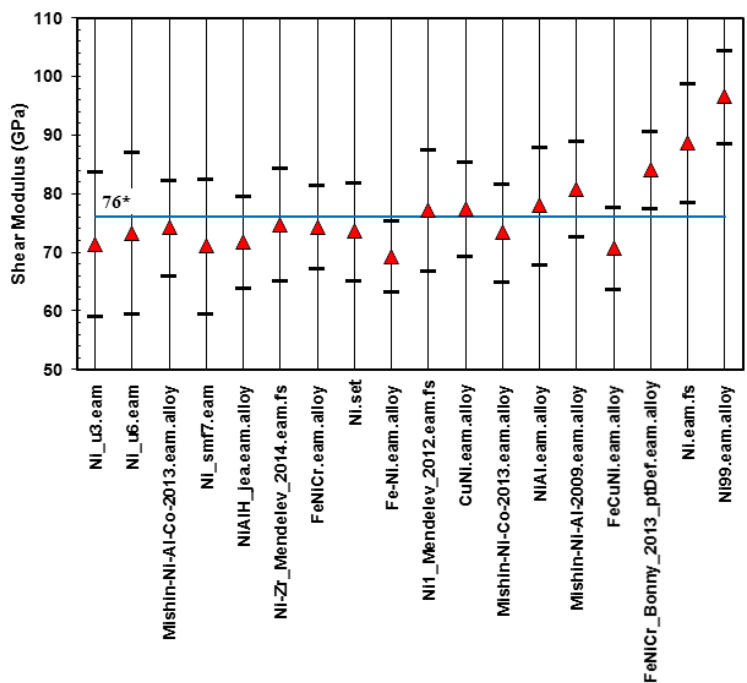

Fig. 8 Nickle Shear modulus predicted by VRH approximation using different force fields

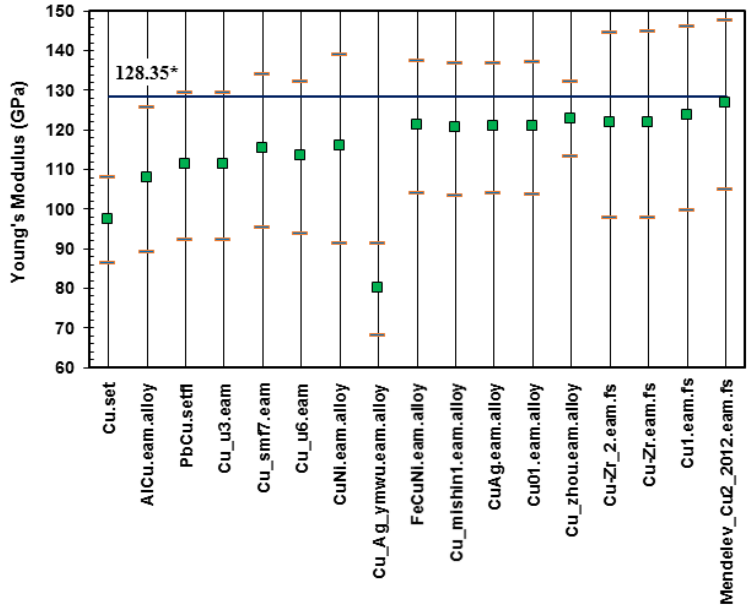

Fig. 9 Copper Young's modulus predicted by VRH approximation using different force fields

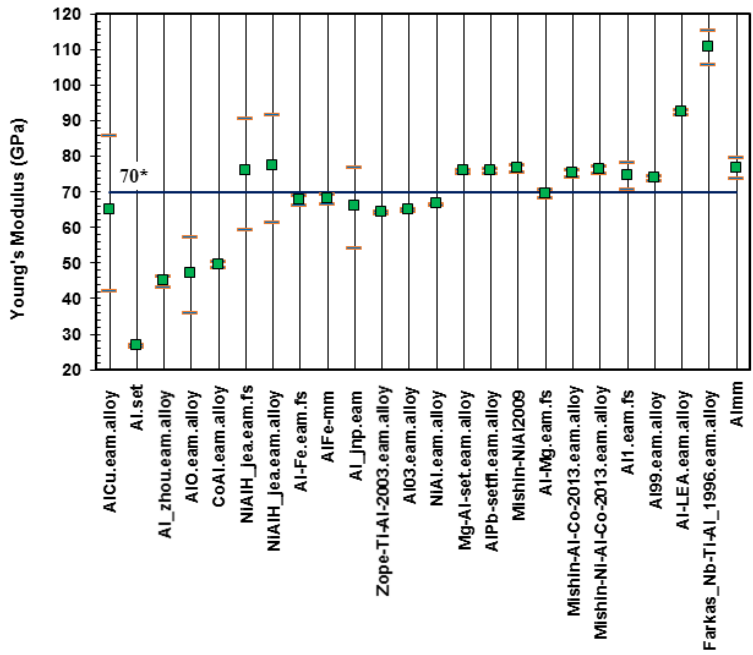

Fig. 10 Aluminum Young's modulus predicted by VRH approximation using different force fields

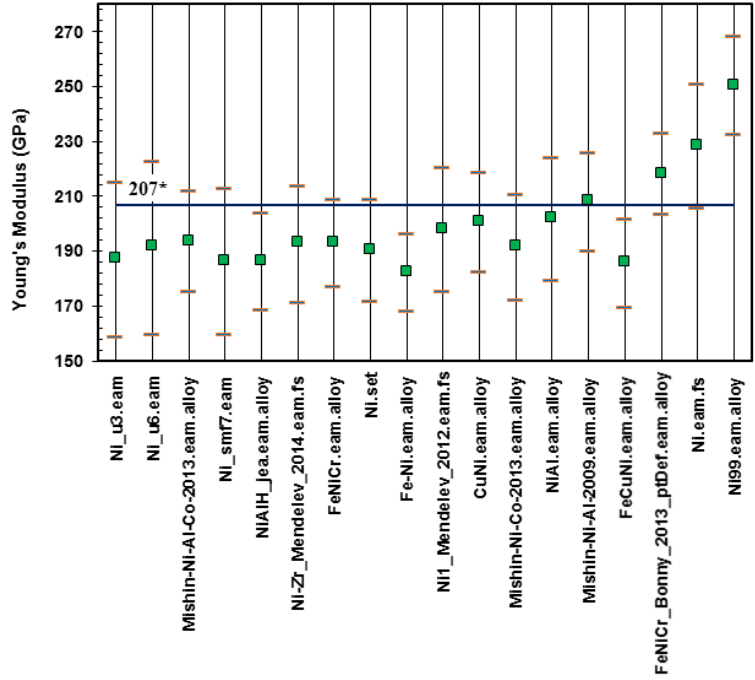

Fig. 11 Nickle Young's modulus predicted by VRH approximation using different force fields

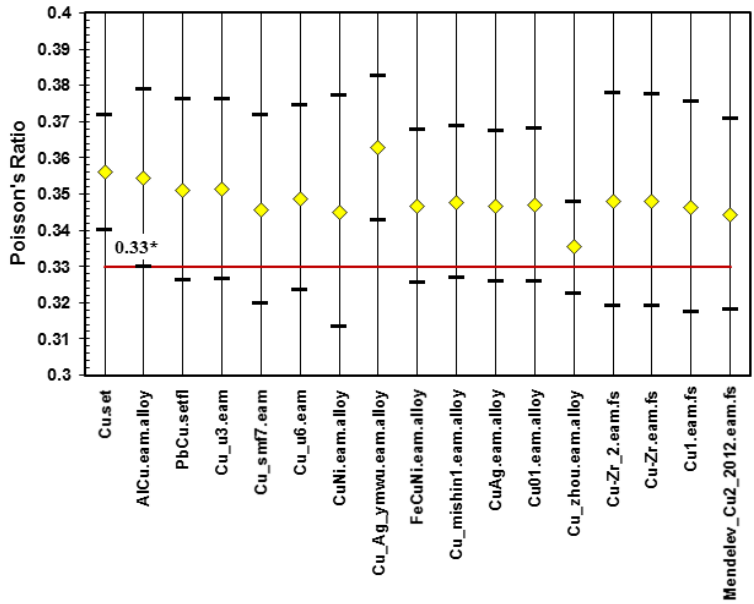

Fig. 12 Copper Poisson's ratio predicted by VRH approximation using different force fields

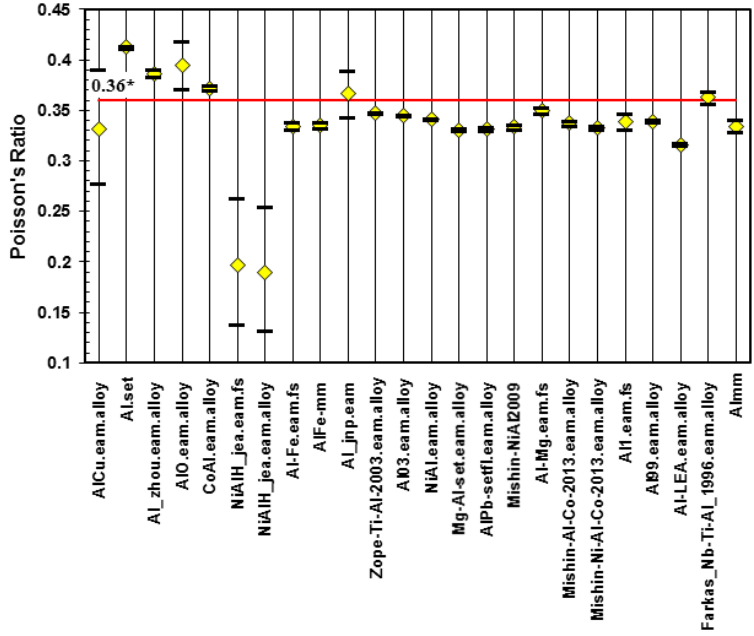

Fig. 13 Aluminum Poisson's ratio predicted by VRH approximation using different force fields

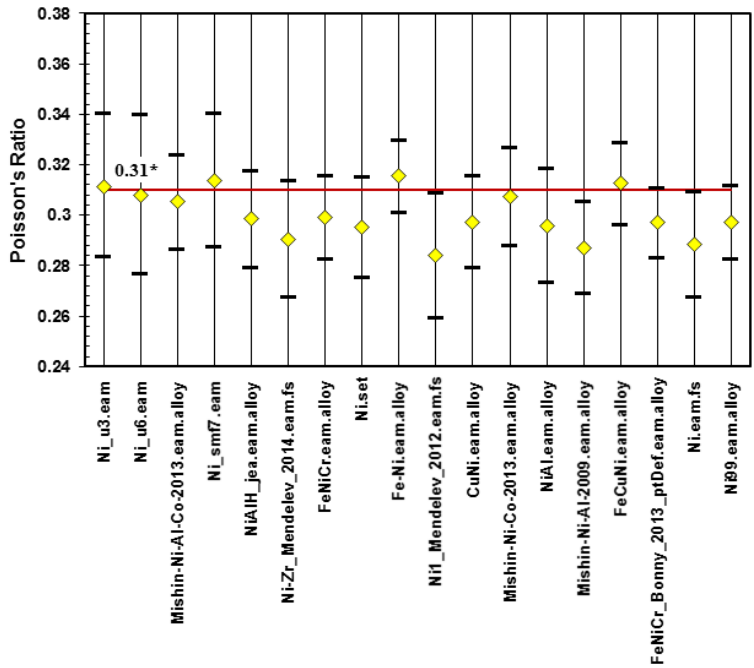

Fig. 14 Nickle Poisson's ratio predicted by VRH approximation using different force fields

elastic stiffness constants $C_{11}$ and $C_{12}$ are very close to the experimental results with the lowest relative errors. Although, it does not mean that other potential are not effective or applicable, users according to their problem and demand may use the other ones which are yielding acceptable errors. Therefore, one may read the original reference of each potential and see for what purpose it has been developed, and according to what reference data they were fitted. To confirm this statement, in continue readers will see that other potentials maybe recommended for finding the other elastic properties rather than or further to what mentioned above for each species.

According to Fig. 5, for Nickel, the two *FeNiCr_Bonny_2013_ptDef.eam.alloy* and *Ni.eam.fs* predict the Nickel bulk modulus with 2.42% and 1.95% relative error with respect to experimental values, respectively. However most of the examined potentials for Nickel may predict an acceptable value with a relative error around 11% such as: *Ni_u3.eam, Ni_u6.eam, Mishin-Ni-Al-Co-2013.eam.alloy, Ni_smf7.eam, Fe-Ni.eam.alloy, CuNi.eam.alloy, Mishin-Ni-Co-2013.eam.alloy, NiAl.eam.alloy, Mishin-Ni-Al-2009.eam.alloy, FeCuNi.eam.alloy, Ni99.eam.alloy*, but they may not be as powerful as the two mentioned ones in predicting both of the stiffness constants. Figure 5 also confirms this postulate that some of the interatomic potentials which have been created for a compound or alloy system, does not necessarily work well for all the existing elements. The two good examples for this statement are *Mishin-Ni-Al-Co-2013.eam.alloy and Mishin-Ni-Al-2009.eam.alloy* that are very accurate for Al bulk modulus with less than 2.00% while they are not this much accurate for Ni and also its elastic stiffness constants. *FeCuNi.eam.alloy* is another one which is much more accurate for *Cu* than *Ni*.

These elastic moduli for the three elements are displayed in terms of VRH approximation with the Voigt and Reuss bounds shown in Figures 6-14. Figures 6-8 depict the shear modulus obtained for the three species based on elastic stiffness constants via different potentials. According to Fig. 6, the accurate copper potentials introduced for bulk modulus are accurate for finding shear modulus though. The difference is increasing their error from experimental result upto 3-4%. The reason is that, for calculating Shear modulus, Young's modulus and Poisson's ratio of an isotropic polycrystalline, in addition to $C_{11}$ and $C_{12}$, $C_{44}$ plays an important role in predicting these elastic properties. This fact has a significant effect on choice of appropriate potential for predicting shear modulus further to/rather than those recommended for finding bulk moduli since *Al* potentials are found to be more precise in finding $C_{11}$ and $C_{12}$ in compare to $C_{44}$ according to Tables 1. In this regard the role of the some other potentials that can predict shear moduli more accurately with an acceptable error for the three elastic constants comes in. For aluminum, it can be obtained from Fig. 7 that *NiAl.eam.alloy, Al-Mg.eam.fs, , Zope-Ti-Al-2003.eam.alloy, Al03.eam.alloy, Al-Fe.eam.fs and AlFe-mm* can predict the *Al* shear moduli better than the others with less than 4.00% relative error and are recommended respectively according to their capability of predicting both shear modulus and elastic stiffness constants based on Table. 2.

Simulation results for nickel shear modulus are demonstrated in Fig. 8 which clearly shows that most of the nickel interatomic potentials, in addition to those mentioned as accurate ones for predicting bulk moduli, predict the shear moduli precisely. This is due to the fact that some nickel interatomic potentials are more accurate in calculating $C_{44}$ in compare to $C_{11}$ and $C_{12}$. Subsequently, accurate $C_{44}$ as an important element in calculating shear moduli, compensate the $C_{11}$ and/or $C_{12}$ with greater errors for some cases. *CuNi.eam.alloy, NiAl.eam.alloy* and *Mishin-Ni-Al-2009.eam.alloy* are instances of this type of force fields.

The above mentioned considerations for predicting shear modulus of the three pure species are applies for Young's modulus and Poisson's ratio as well which can be seen through Figs. 9-14. From Figs. 3-14, it is clear that, according to our expectation, interatomic potentials that accurately predicted all the three elastic constant with a reasonable error are correctly yielding Shear moduli, Young's moduli and Poisson's ratio of copper, aluminum and nickel. As mentioned earlier, some potentials maybe useful for obtaining one or two of the three independent elastic constants and also elastic modulus. Thus, users should be aware of the mentioned deceptive results. They are not too many for bulk moduli while it is expected to be more in finding of other elastic modulus because of another variable ($C_{44}$) which merges into an algebraic expression and make the scenario much more complicated. Therefore, it is highly recommended that users pay attention to the accuracy of the elastic constants which are also able to estimate an accurate isotropic elastic moduli.

## Conclusion

Adequate choice of interatomic potential, as frequent concern in conversations among the molecular dynamic simulation community, was studied for finding the three independent elastic stiffness constants, the $C_{11}$, $C_{12}$ and $C_{44}$, of copper, aluminum and nickel cubic single crystals.. The three independent stiffness constants of cubic single crystals originated from implementation of all the studied potentials, provided from NIST IPR and LAMMPS database, were determined using a linear regression analysis of the linear part of stress-strain curves. Once elastic constants were obtained, they have been converted into isotropic elastic moduli using Voigt-Reuss-Hill approximation. Bulk modulus, shear modulus, Young's modulus and Poisson's ratio were calculated and results from each interatomic potential have been compared to the experimental ones both for anisotropic elastic stiffness constants and the converted isotropic elastic modulus. It is concluded that inadequate choice of force field strongly affects the simulation results and gives rise to some inconveniences for calculations. Some of the interatomic potentials seem to be useful and accurate for predicting one or two of the elastic constants or elastic modulus, not all of them. It is also found that the elemental potentials that have been generated for a specific alloy or compound, is not expected to necessarily work for all of the present species in the compound. Results presented in this work are useful for interested researcher in the field of atomistic study of materials mechanical properties and increase the assurance of the users to see which interatomic potential fits well to their specific problem.

**Acknowledgment**
Support for this work by the U.S. National Science Foundation under grant number CBET-1404482 is gratefully acknowledged.